\documentclass[preprint,amsmath,amssymb]{elsarticle}
\usepackage{amssymb}
\usepackage{graphicx}
\usepackage{subfigure}
\usepackage{dcolumn}
\usepackage{bm}
\usepackage{hyperref}

\begin{document}

\begin{frontmatter}

\title{Unified description of perturbation theory and band center anomaly
in one-dimensional Anderson localization}

\author{Kai Kang}
\ead{colinkk@pku.edu.cn}
\address{School of Physics, Peking University, Beijing 100871, China}
\author{Shaojing Qin}
\address{Institute of Theoretical Physics, Chinese Academy of
Sciences, P.O. Box 2735, Beijing 100190, China}
\author{Chuilin Wang}
\address{China Center of Advanced Science and Technology,
P. 0. Box 8730, Beijing 100190, China}

\date{\today}
\begin{abstract}
We calculated numerically the localization length of one-dimensional
Anderson model with diagonal disorder. For weak disorder, we showed
that the localization length changes continuously as the energy
changes from the band center to the boundary of the anomalous region
near the band edge. We found that all the localization lengths for
different disorder strengths and different energies  collapse onto a
single curve, which can be fitted by a simple equation. Thus the
description of the perturbation theory and the band center anomaly
were unified into this equation.
\end{abstract}


\end{frontmatter}

\section{Introduction}
\label{sec:intro}

Electronic transport properties, the motion of electrons, in a
random potential are closely related to the phenomenon of
localization. Since the pioneering work of Anderson
\cite{PAnderson1958} for disordered systems fifty years ago,
Anderson localization has been applied to various fields including
photonics and cold atoms \cite{AdLagendijk2009}. However, even for
the simplest one-dimensional case, we have not found exact
analytical treatment for arbitrary disorder strengths and energies
\cite{bkramer1993,SRyu2004,FEvers2008}. Accurate numerical
approaches have been developed by using the quantum transfer matrix
renormalization group method for finite temperature systems
\cite{lpyang2009}, the density matrix renormalization group method
for interacting systems \cite{pshmitteckert1998}, and the integral
equation method for systems in the thermodynamic limit
\cite{kkang2010}, respectively. In this work we will give a unified
description of the in-band (without the band edge anomalous region)
behavior for the one-dimensional model with weak diagonal disorder.
We study the problem with the help of a numerical method we
developed earlier in Ref.~\cite{kkang2010}, which was an application
of the transfer matrix method \cite{jbpendry1994} in localized phase
in the thermodynamic limit.

The localization theory  shows rigorously that all the eigenstates
are exponentially localized for one-dimensional uncorrelated
disordered systems. The single parameter scaling (SPS) theory
\cite{EAbrahams1979} gave great insight into the properties of
disordered systems. It argues that the dimensionless conductance $g$
is the only relevant parameter that controls its variation with
system size $L$. SPS requires that the complete distribution
function $p(g)$ should be determined by one single parameter.
Originally, SPS was derived within the random phase approximation
\cite{DFisher1980}. The finite size Lyapunov exponent
$\tilde{\gamma}(L)=(1/2L)\ln(1+1/g)$ was proposed to be an
appropriate scaling parameter (it approaches the nonrandom limit
when $L\rightarrow\infty$ \cite{ILifshitz1988}). It is well-known
that the random phase approximation fails at the band edges $E=\pm2$
as well as the band center $E=0$ \cite{HSchomerus2002,AStone1983}.
Another length scale was suggested to give an appropriate criterion
for the failure of SPS \cite{LDeych2001} and a relatively good
scaling parameter was found for the non-SPS region
\cite{ALisyansky2003}.

For weak disorder, although $p(g)$ has two independent parameters
which is in contrast with the case of strong disorder, it was found
that one of them is a universal number \cite{PLee1985}. Thus it is
still SPS. There exists a perturbation theory in which the Lyapunov
exponent depends on energy $E$ and disorder strength $W$
\cite{DThouless1979}
\begin{equation}
\gamma_p(E,W^2)=\frac{W^2}{96(1-E^2/4)}.\label{eq:sp}
\end{equation}
At the band center, a revised perturbation gives
\cite{MKappus1981,BDerrida1984,FIzrailev1998}
\begin{equation}
\gamma(W^2)=\frac{W^2}{105.045\cdots}.\label{eq:c}
\end{equation}
There was no analytical expression which connects smoothly the above
two perturbative results. To clarify the smooth crossover in between
we study the Lyapunov exponent with two variables, the energy $E$
and the disorder strength $\sigma^2$ in this paper ($\sigma^2
=W^2/12$). Furthermore, we only study the problem in the
thermodynamic limit and thus the third variable, the size of a
finite system, is not in consideration.

It was shown that for one-dimensional Anderson model with diagonal
disorder, the band center is actually a band boundary adjoining two
neighbor bands rather than the center of a single band, which is the
reason for the fail of SPS near the band center. Therefore it is
similar to the band edge \cite{BAltshuler2003}. This should also be
why in Ref.~\cite{FIzrailev1998} the authors used a four-step map at
the band center. In our recent work, we proposed a parametrization
method of the transfer matrix for one-dimensional Anderson model
with diagonal disorder. Making use of this method, we obtained
numerical results of the dependence of the Lyapunov exponent on
energy and disorder strength in the thermodynamic limit within the
localization regime. A remarkable coincidence of our numerical
results with the existed results demonstrated the reliability and
efficiency of our method \cite{kkang2010}. For the anomalous region
near the band center, there have been analytical results for
$E\rightarrow0$ and $E\gg\sigma^2$ through the scaled energy
$E/\sigma^2$ \cite{BDerrida1984}. In the present work, we will show
the detailed numerical results about the anomalous behavior of the
Lyapunov exponent with weak disorder near the band center by using
our method, which coincides with the known analytical results; and
we will give a simple equation to describe the behavior for
arbitrary weak disorder strength and for all $E$ in the band except
the anomalous region near the band edge through the parameter
$E/\sigma^2$. Thus the band center anomaly and the perturbation
theory are connected smoothly.

\section{Parametrization method}

Consider the one-dimensional Anderson model with diagonal disorder,
\begin{equation}
\psi_{i-1}+\psi_{i+1}=(E-\epsilon_i)\psi_i, \label{eq:se}
\end{equation}
where $\psi_i$ is the electron wavefunction at site-$i$ and
$\epsilon_i$ is the on-site energy from a random distribution
$p_\epsilon(\epsilon)$. For the box distribution,
$p_\epsilon(\epsilon)=1/W$, $|\epsilon|<W/2$,
$\langle\epsilon^2\rangle=W^2/12$, and for the Gaussian
distribution,
$p_\epsilon(\epsilon)=(1/\sqrt{2\pi}\sigma)\exp(-\epsilon^2/2\sigma^2)$,
$\langle\epsilon^2\rangle=\sigma^2$. In the transfer matrix method,
Eq.~(\ref{eq:se}) can be written as
\begin{equation}
\Psi_{i+1}=\left(\begin{array}{c} \psi_{i+1} \\ \psi_i
\end{array}\right)= \left(\begin{array}{cc} v_i & -1 \\ 1 & 0
\end{array}\right) \left(\begin{array}{c} \psi_i \\ \psi_{i-1}
\end{array}\right)=\mathbf{T}_i\Psi_i, \label{eq:tm}
\end{equation}
where $v_i=E-\epsilon_i$ and $\mathbf{T}_i$ is the transfer matrix.

Using a parametrization method of the transfer matrix proposed in
our previous work \cite{kkang2010}, we can calculate the Lyapunov
exponent in the thermodynamic limit within the localization regime.
Let $\mathbf{M}_L=\mathbf{T}_L\mathbf{T}_{L-1}\cdots\mathbf{T}_1$.
Then we can parameterize $\mathbf{M}\mathbf{M}^t$ as follows
\begin{equation}
\mathbf{U}(\theta_L)\mathbf{M}_L\mathbf{M}^t_L \mathbf{U}(-\theta_L)
=\left(\begin{array}{cc} e^{\lambda_L} & \\ & e^{-\lambda_L}
\end{array} \right),
\end{equation}
where $\mathbf{M}^t$ is the transpose of $\mathbf{M}$ and
\begin{equation}
\mathbf{U}(\theta_L)=\left(\begin{array}{cc} \cos\theta_L & -\sin\theta_L
\\ \sin\theta_L & \cos\theta_L \end{array} \right).
\end{equation}
The recursion relation of $\theta$ in large $L$ limit is
\begin{equation}
\tan\theta_{L+1}=\frac{1}{v_{L+1}-\tan\theta_L}. \label{eq:theta}
\end{equation}
The two parameters $\lambda$ and $\theta$ are what we need in order
to calculate the Lyapunov exponent $\gamma$. The equations we
obtained for the distribution function $p(\theta)$ and the Lyapunov
exponent are
\begin{equation}
p(\theta)=\frac{1}{\sin^2\theta}\int p(\theta')p_v(
\frac{1}{\tan\theta}+\tan\theta')\mathrm{d}\theta'
\label{eq:theta_distri}
\end{equation}
\begin{equation}
\gamma=-\int p(\theta)\ln |\tan \theta |\mathrm{d}\theta
\label{eq:length}
\end{equation}
where $p_v(v)$ is the  random distribution function of $v$. In
this study we use the Gaussian distribution
$p_v(v)=(1/\sqrt{2\pi}\sigma)\exp[-(v-E)^2/2\sigma^2]$ to solve
Eq.~(\ref{eq:theta_distri}) and to calculate $\gamma$
numerically for different $\sigma^2$ and different energy $E$.

\section{Numerical results}

\begin{figure}
\includegraphics[width=8cm]{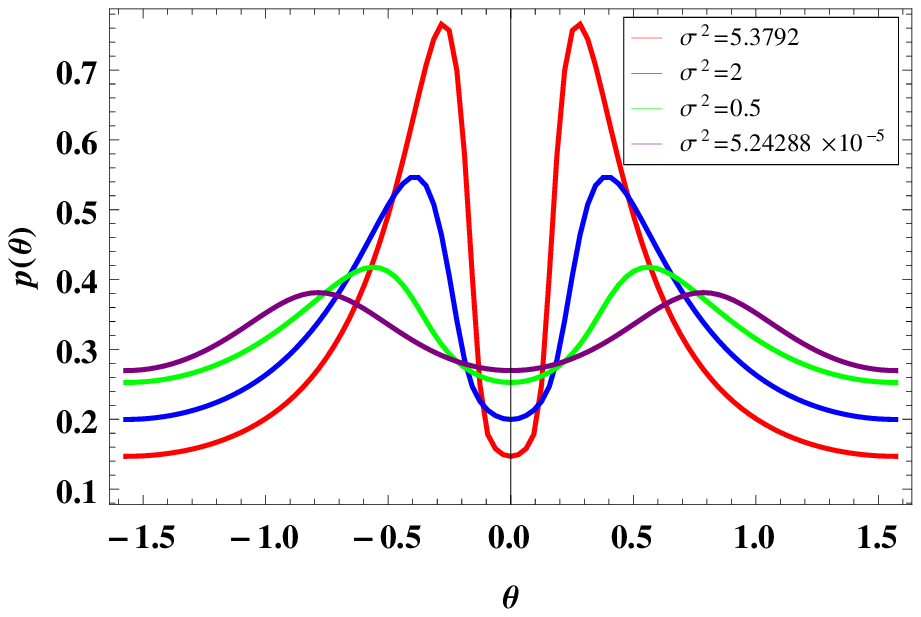}
\caption{(Color online) Distribution $p(\theta)$ from
Eq.~(\ref{eq:theta_distri}) at fixed $E=0$ for the Gaussian
distribution with decreasing $\sigma^2$.} \label{fig:figure1}
\end{figure}
\begin{figure}
\includegraphics[width=8cm]{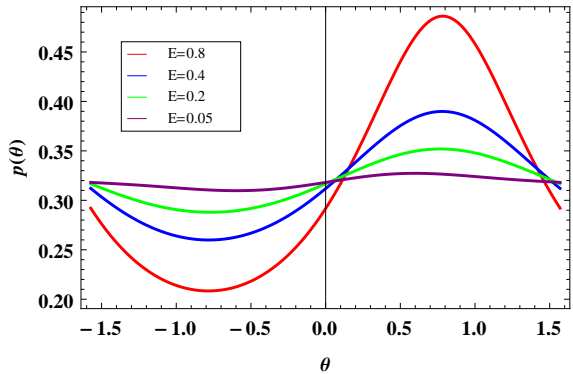}
\caption{(Color online) Distribution $p(\theta)$ from
Eq.~(\ref{eq:theta_distri}) at fixed $\sigma^2=0.00125$ for the
Gaussian distribution with decreasing $E$.} \label{fig:figure2}
\end{figure}
Fig.~\ref{fig:figure1} and Fig.~\ref{fig:figure2} illustrate the
distribution functions $p(\theta)$ which were calculated as $p_v(v)$
being the Gaussian distribution. In Fig.~\ref{fig:figure1}, four
curves of $E=0$ are plotted for different values of $\sigma^2$
decreased from 5.3792 to $5.24288\times10^{-5}$; and in
Fig.~\ref{fig:figure2}, four curves of $\sigma^2=0.00125$ are
plotted for different values of $E$ decreased from 0.8 to 0.05. In
Fig.~\ref{fig:figure1}, $p(\theta)$ becomes a stationary
distribution when disorder is weak enough (the line of
$\sigma^2=5.24288\times10^{-5}$)
\begin{equation}
p(\theta)=\frac{1}{K(1/2)\sqrt{3+\cos4\theta}},
\end{equation}
where $K$ is the complete elliptic integral of the first kind
\cite{FIzrailev1998}. There is a correspondence between $p(\theta)$
and the distribution of the reflection phase $\phi$ in
Ref.~\cite{CBarnes1990}, which gave similar plots, through the
relation $\tan\theta=\cos(\phi/2)/\cos(\mu+\phi/2)$, where
$2\cos\mu=E$. In Fig.~\ref{fig:figure2}, the expression of
distribution $p(\theta)$ for the three larger values of $E$ is
\begin{equation}
p(\theta)=\frac{\sin\mu}{\pi(1-\cos\mu\sin2\theta)}, \label{eq:uni}
\end{equation}
corresponding to the uniform distribution for the perturbation
theory \cite{MKappus1981,FIzrailev1998}. A recent numerical
calculation \cite{tkaya2009} of the angle distribution function in
Ref.~\cite{FIzrailev1998} also gave a similar behavior. In our
numerical approach we have set a relative precision $10^{-10}$ of
$p(\theta)$.

\begin{figure}
\includegraphics[width=8cm]{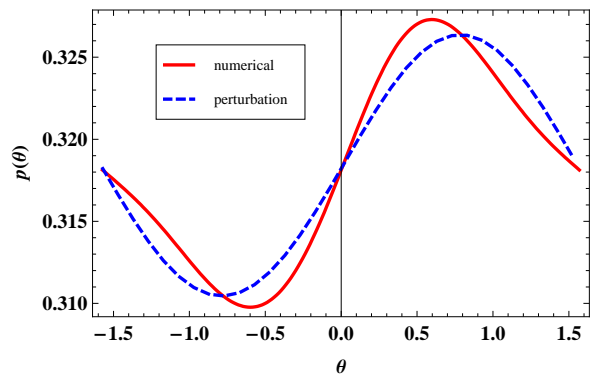}
\caption{(Color online) Comparison of $p(\theta)$  between the
numerical result and Eq.~(\ref{eq:uni}) ($E=0.05$ and $\sigma^2=0.00125$), which means that the
standard perturbation theory fails when $E$ approaches $\sigma$.}
\label{fig:figure3}
\end{figure}

For weak disorder we have known that the perturbation theory breaks
down near the band center. It is clear in Fig.~\ref{fig:figure3}
that the distribution $p(\theta)$ of $E=0.05$ with
$\sigma^2=0.00125$ deviates from Eq.~(\ref{eq:uni}), which means
that the band center anomalous region extends up to $E>80\sigma^2$.
In Figs.~\ref{fig:figure1} and \ref{fig:figure2} we have
demonstrated known results for the two limits $E/\sigma^2
\rightarrow0$ and $E/\sigma^2 \rightarrow \infty$; and in
Fig.~\ref{fig:figure3} contribution of higher order terms to the
curves in weak disorder perturbation becomes apparent for finite
$E/\sigma^2$. In Ref.~\cite{HSchomerus2003} the authors gave a
complete accurate analysis of not only the localization length, but
all higher moments of the distribution of the Lyapunov exponent for
finite systems. It is natural to expect that the deviation from
Eq.~(\ref{eq:uni}) comes from higher order terms in the perturbation
given in Ref.~\cite{HSchomerus2003}. And the present work will give
an accurate equation for the localization length in the whole band
except the band edge anomalous region.

\begin{figure}
\includegraphics[width=8cm]{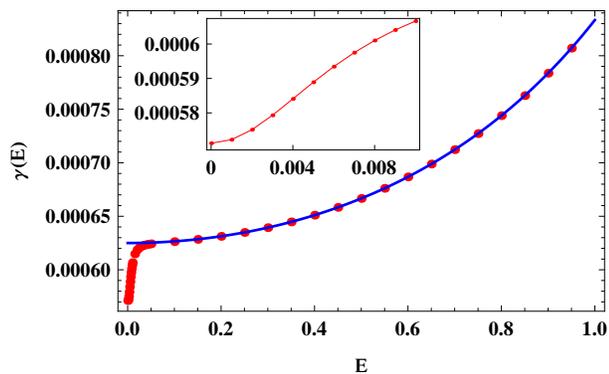}
\caption{(Color online) $\gamma(E)$ with disorder strength
$\sigma^2=0.005$ for $E<1$ and $p_v(v)$ as the Gaussian
distribution. The solid line is the standard perturbation result.
A magnified curve for $E<0.01$ is plotted in the inset.} \label{fig:figure4}
\end{figure}

To demonstrate the anomalous behavior near the band center, we plot
$\gamma(E)$ with disorder strength $\sigma^2=0.005$ for $p_v(v)$
being the Gaussian distribution in Fig.~\ref{fig:figure4}. It shows
clearly that in the vicinity of the band center the behavior of the
Lyapunov exponent deviates from the perturbation result ($\gamma_p$
solid line) and when $E$ approaches zero it becomes the result of
Eq.~(\ref{eq:c}) smoothly \cite{HSchomerus2003}. Our numerical
results coincides with the analytical results for the region near
$E/\sigma^2\ll1$ and $E/\sigma^2\gg1$ \cite{BDerrida1984}. We will
present a quantitative description by collapsing the curves for
different energies and disorder strengths through the parameter
$E/\sigma^2$.

\begin{figure}
\includegraphics[width=8cm]{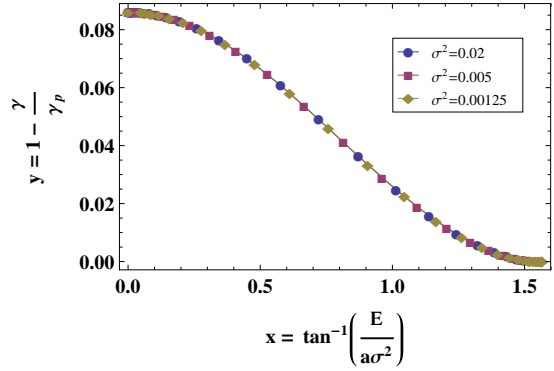}
\caption{(Color online) Three groups of data fall on a single line
with our choice of $y$ and $x$. Eq.(\ref{eq:fit}) coincides well
with the numerical data in this figure} \label{fig:figure5}
\end{figure}

Considering Eqs.~(\ref{eq:sp}) and (\ref{eq:c}) as well as the
result shown in Fig.~\ref{fig:figure4}, we choose
$y=1-\gamma/\gamma_p$ as a scaling parameter. $y$ is the relative
difference between the perturbation result and the numerical result.
To show all the values of $E/\sigma^2$ we use
$x=\arctan(E/a\sigma^2)$, where $a$ is a fitting coefficient. $x=0$
at the band center $E=0$ and $x=\pi/2$ for a finite energy when
$E/\sigma^2\sim1/\sigma^2\rightarrow\infty$ with
$\sigma^2\rightarrow0$. In Fig.~\ref{fig:figure5} there are three
groups of data and the fitting coefficient $a\approx1.436$. Each
group has a fixed weak disorder strength. All the data points merge
into a single curve. Thus the in-band behavior can be described by a
single scaling parameter for all weak disorder values of $\sigma^2$
as long as $E$ is not located in the band edge anomalous region. If
the energy is in the band edge anomalous region, we found that $y$
deviates from the universal curve in Fig.~\ref{fig:figure5}, which
was not shown here.

When $E$ approaches the boundary of the band center anomalous
region, $\gamma$ becomes equal to the perturbation result
$\gamma_p$, therefore $y\rightarrow0$; when $E=0$,
$y=1-96/105.045\approx0.086$. So We expect that
\begin{equation}
y\approx(1-\frac{96}{105.045})\cos^2x \label{eq:fit},
\end{equation}
where $\cos^2x=[1+(E/a\sigma^2)^2]^{-1}$. Therefore the Lyapunov
exponent for weak disorder and with energy in the band except near
the band edge is
\begin{equation}
\gamma=f(\frac{E}{\sigma^2})\gamma_p, \label{eq:quantit}
\end{equation}
where $f(E/\sigma^2)=1-0.086[1+(E/a\sigma^2)^2]^{-1}+\cdots$. The
difference in $\gamma$ is only $8.6\%$ for the $E=0$ and a finite
energy.  A quantitative description can be given: if we set the
criterion for the breaking down of the perturbative result in
Eq.~(\ref{eq:sp}) as $1\%$ difference in $\gamma$, then by
Eq.~(\ref{eq:quantit}) we obtain $|E|\gtrsim4\sigma^2$  as the
region for Eq.~(\ref{eq:sp}) to be considered as valid. It is
$|E|\gtrsim8D$ which is in agreement with the $|E|\gtrsim10D$ given
in Ref.~\cite{HSchomerus2003}. Therefore there is no definite
boundary between the perturbation theory region and the band center
anomalous region.

\begin{figure}
\includegraphics[width=8cm]{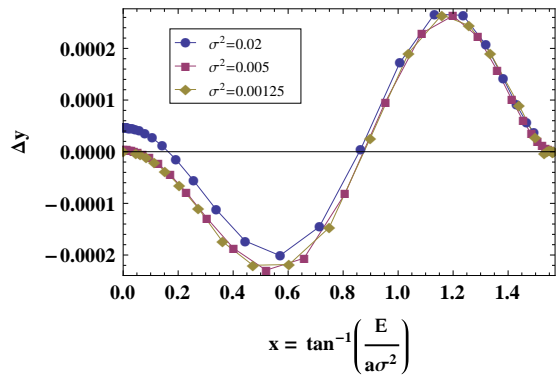}
\caption{(Color online) The difference between the numerical data
and Eq.~(\ref{eq:fit}).} \label{fig:figure6}
\end{figure}

In Fig.~\ref{fig:figure5} all points collapse onto a single curve.
We should mention that Eq.~(\ref{eq:fit}) only includes the main
correction from $E/\sigma^2$. There are higher order corrections in
$f(E/\sigma^2)$ \cite{MKappus1981}. The small deviation between
Eq.~(\ref{eq:fit}) and the numerical data was shown in
Fig.~\ref{fig:figure6}.

At $x=0$ point, where $E=0$, it is already known that
Eq.~(\ref{eq:c}) is only valid in weak disorder limit. $\gamma$
behaves in logarithmic of the disorder strength in strong disorder
limit, and the crossover curve between the two limits can be found
in Ref.~\cite{kkang2010}. The difference around $x=0$ point is
reasonable: $\sigma^2=0.02$ is not weak enough and higher order
corrections of the perturbation in $\sigma^2$ becomes apparent. In
the weak disorder limit ($\sigma^2=0.005,0.00125$ in
Fig.~\ref{fig:figure6}) all the curves for different $\sigma^2$ is a
single curve, which depends on $E/\sigma^2$ only.

\section{Conclusion}

In summary, we calculated the inverse localization length in
one-dimensional Anderson model with diagonal disorder. We obtained
numerically the curve of the inverse localization length for
energies inside the energy band in the case of weak disorder. A
unifying curve was given for different weak disorder strengths and
energies outside the band edge anomalous region.

The scaled energy used to plot the unifying curve is $E/\sigma^2$,
which has correspondence to the scaling parameter
$\kappa=l_{loc}/l_s$ proposed by Deych et al.~\cite{BAltshuler2003}.
They used $\kappa$ for the criterion of SPS, which has a clear
physical image. However, the new length scale $l_s$ and the
localization length $l_{loc}$ are both functions of the energy $E$
and the disorder strength $\sigma^2$. Thus $\kappa$ is a two
parameter function $\kappa(E,\sigma^2)$. And we showed in this work
that $\kappa$ actually depends on the single parameter $E/\sigma^2$,
which is the essential parameter to give the universal curve in
Fig.~\ref{fig:figure5}.

We also found a similar unifying description of the perturbation
theory and the band center anomaly for the box distribution of
disorder. We believe that other kinds of uncorrelated disorder with
a finite variance can have the same unifying description. It should
be mentioned that the Lorentzian distribution has no such band
center anomaly, because its variance does not exist.

\section{Acknowledgments}

This work was supported by National Natural Science Foundation of
China No.~10374093, the National Program for Basic Research of MOST
of China, and the Knowledge Innovation Project of Chinese Academy of
Sciences.

\bibliographystyle{elsarticle-num}
\bibliography{bib}

\end{document}